\documentclass[sigconf,authorversion,nonacm]{acmart}

\usepackage{booktabs} %

\usepackage{algorithm}
\usepackage{algpseudocode,mathrsfs}
\usepackage{multirow}
\usepackage{graphicx}
\usepackage{caption}
\usepackage{subcaption}
\newcommand{\T}{^\mathsf{T}}
\newcommand{\inv}{^\mathsf{-1}}

\newcommand{\hatH}{\hat{H}}

\newcommand{\C}{\mathcal{C}}
\newcommand{\Beta}{\mathcal{B}}

\newcommand{\hatg}{\hat{g}}

\newcommand{\checkBeta}{\check{\Beta}}

\newcommand{\checkQ}{\check{Q}}

\newcommand{\dx}{\delta x}
\definecolor{awesome}{rgb}{1.0, 0.13, 0.32}

\copyrightyear{2023}
\acmYear{2023}
\setcopyright{acmlicensed}\acmConference[SA Technical Communications
'23]{SIGGRAPH Asia 2023 Technical Communications}{December 12--15,
2023}{Sydney, NSW, Australia}
\acmBooktitle{SIGGRAPH Asia 2023 Technical Communications (SA Technical
Communications '23), December 12--15, 2023, Sydney, NSW, Australia}
\acmPrice{15.00}
\acmDOI{10.1145/3610543.3626161}
\acmISBN{979-8-4007-0314-0/23/12}

\begin{document}
\title{Efficient Incremental Potential Contact for Actuated Face Simulation}
\author{Bo Li}
\authornote{Contributed equally to this work.}
\orcid{0009-0009-8444-7718}
\affiliation{\institution{ETH Zurich}\country{Switzerland}}
\email{bolibo@ethz.ch}
\author{Lingchen Yang}
\authornotemark[1]
\orcid{0000-0001-9918-8055}
\affiliation{\institution{ETH Zurich}\country{Switzerland}}
\email{lingchen.yang@inf.ethz.ch}
\author{Barbara Solenthaler}
\orcid{0000-0001-7494-8660}
\affiliation{\institution{ETH Zurich}\country{Switzerland}}
\email{solenthaler@inf.ethz.ch}

\renewcommand\shortauthors{B. Li, L. Yang, B. Solenthaler}

\begin{teaserfigure}
\centering
  \includegraphics[width=1.0\textwidth]{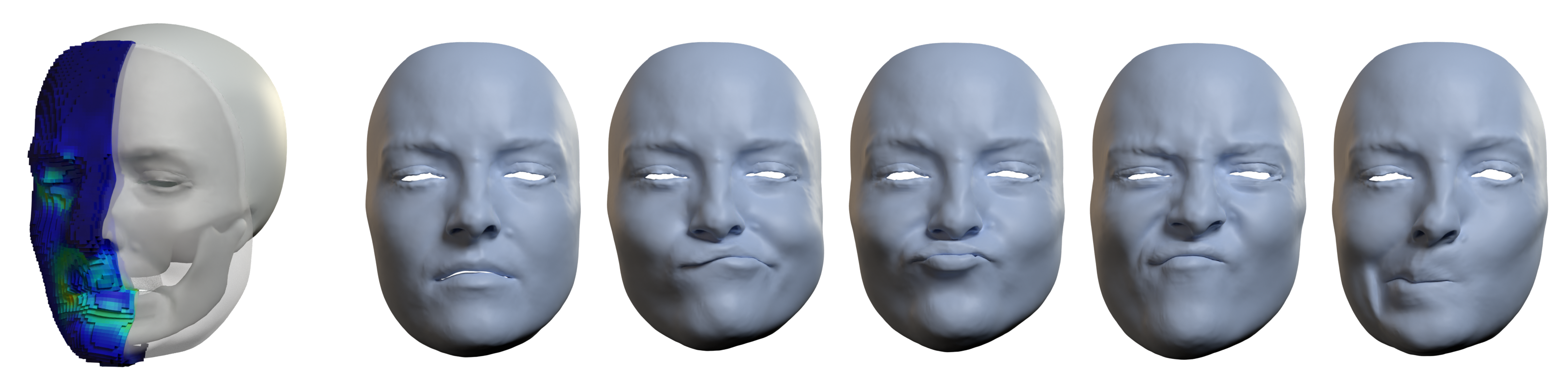}
  \caption{We model the human face as a soft body and simulate its deformation using Finite Element Method driven by the actuation of muscles. With the introduction of collision handling, we are capable of simulating challenging expressions that may result in unrealistic self-intersections if not solved properly. }
  \label{fig:teaser}
\end{teaserfigure}

\begin{abstract}
We present a quasi-static finite element simulator for human face animation. We model the face as an actuated soft body, which can be efficiently simulated using Projective Dynamics (PD). We adopt Incremental Potential Contact (IPC) to handle self-intersection. However, directly integrating IPC into the simulation would impede the high efficiency of the PD solver, since the stiffness matrix in the global step is no longer constant and cannot be pre-factorized. We notice that the actual number of vertices affected by the collision is only a small fraction of the whole model, and by utilizing this fact we effectively decrease the scale of the linear system to be solved. With the proposed optimization method for collision, we achieve high visual fidelity at a relatively low performance overhead.

\end{abstract}

\ccsdesc[500]{Computing methodologies~Physical simulation}

\keywords{Physically based simulation, Projective Dynamics, Incremental Potential Contact, Collision Handling}

\maketitle

\enlargethispage{25pt}

\section{Introduction}

Various deformable entities, such as human facial structures, can be viewed as soft bodies whose motion is governed by the underlying muscle activation. Yang et al. \cite{Lingchen2022Fem} introduced a neural model that implicitly represents muscle actuation as a function of spatial coordinates, which drives the soft body simulation using Projective Dynamics (PD) \cite{bouaziz2014pd}. Yet, this simulation framework does not account for collisions, an aspect which is crucial to realistic animation. For example, contact is ubiquitous around the mouth (see Figure~\ref{fig:teaser}, right part), and the inter-penetration between the upper and lower lips would generate implausible artifacts.

Recently, Incremental Potential Contact (IPC) \cite{li2020ipc} has gained much popularity for contact modeling. It could achieve large-scale intersection-free deformations. In this work, we integrate IPC into the simulator by Yang et al. Although the integration may seem simple at first glance, there are inherent incompatibilities between IPC and PD. The efficiency of PD is inherited from its reuse of the pre-factorized stiffness matrix for solving a fixed linear system, but a straightforward introduction of the time-varying Hessian matrix of IPC could negate this benefit.

The original collision handling method in the PD framework projects penetrating vertices onto the closest surface and employs spring-like energy terms to redirect them to the collision-free state. However, in addition to slight penetration, it suffers from sticking artifacts since the system only generates repulsion when the penetration really happens. To address this issue, Lan et al. \cite{lan2022penetration} proposed a "barrier projection" method that strategically sets the target position for collision vertices. It models the target position of a vertex after rebounding from the barrier by utilizing its velocity information. However, this approach is unsuitable for quasi-static simulations where the concept of velocity does not exist, leading to a reversion to the standard PD method. On the other spectrum, Wang et al. \cite{wang2021local} pursued this traditional method, but operated under the premise that collisions predominantly occur within confined areas. This assumption allows them to alleviate the complexity of the linear solving problem by transforming the system from a large, sparse matrix into a smaller, denser one. 

Inspired by the above discussed previous works, we propose to use IPC for efficient collision handling in actuated face simulation. Furthermore, we notice that even a smaller number of the vertices in the confined region are involved in the collision, providing further room for optimization beyond the idea of Wang et al. \shortcite{wang2021local}. To this end, we are able to perform high-fidelity intersection-free simulation with optimized efficiency.
\enlargethispage{20pt}
\section{Simulation Procedure}
\subsection{Actuated Soft Body Simulation}
Unlike passive objects, the motion of muscles is driven by internal activation signals. To describe this mechanism, we can formulate the desired deformation as a symmetric actuation matrix \cite{ichim2017phace, klar2020shape, Lingchen2022Fem}. For the simulation mesh discretized into finite elements and nodal points, this shape-targeting model provides the following elastic potential for each element (denoted by subscript $i$):
\begin{equation}
\label{eq:local}
  E_i(x) = \min_{R_i \in \mathrm{SO}(3)} \|{F_i(x) - R_iA_i}\|^2_F = \|{G_i x - p_i}\|^2,
\end{equation}
where $x \in \mathbb{R}^{3n}$ is the stacked vector of the nodal vertices. The energy measures the distance between the deformation gradient $F_i$ and the actuation matrix $A_i$ using Frobenius norm. $R_i$ is allowed to factorize out their rotational difference. Equivalently, the flattened deformation gradient is written as $G_i x$ where $G_i$ is a gradient mapping matrix, and we use $p_i$ to denote the flattened vector of $R_i A_i$. In the left part of Figure~\ref{fig:teaser}, the simulation mesh is colored based on the strength of actuation. On the other hand, the surface mesh of the face (the transparent layer) is \textit{embedded} inside the simulation mesh via interpolation and deformed accordingly. 

The quasi-static simulation aims to find the equilibrium state with minimal energy. This optimization problem could be effectively solved using Projective Dynamics (PD) \cite{bouaziz2014pd} in a local-global alternating manner. In the \textit{local step}, we find the best rotation $R_i$ using the polar decomposition of $F_i A_i$. This step could be parallelized among finite elements.

Then in the \textit{global step}, $\{R_i\}$ are fixed. The total energy $E(x) = \sum_i w_i E_i(x)$ is simply the sum of all element energies, weighted by the volume $w_i$ at rest state. By setting $\nabla E(x)=0$ we get:
\begin{equation} \label{eq:global}
  \underbrace{(\sum_i w_i G_i\T G_i)}_H x = \sum_i w_i G_i\T p_i .
\end{equation}
The efficiency of PD comes from the fact that the left side $H$ in the global step only depends on the configuration of the FEM system, which remains constant regardless of the actuation state. Thus, its Cholesky factorization could be pre-computed. At runtime, Equation~\ref{eq:global} could be effectively solved by forward/backward substitution.

\subsection{Collision Handling} \label{sec:ipc}
The actuation by itself could drive the face into implausible states with self-intersection. To achieve visually convincing results, we adopt the method of Incremental Potential Contact (IPC)~\cite{li2020ipc} to handle collisions. For each primitive pair (vertex-triangle or edge-edge), their unsigned distance $d_k$ is used to compute a barrier function $b(d_k)$ that penalizes penetration. $b(\cdot)$ is logarithm-like so that the response goes to infinity as the distance approaches zero, similar to the idea of the interior point method. The barrier is activated only when the distance is smaller than a distance threshold $d_0$, thus this threshold determines a constraint set $\C$ that contains the active primitive pairs.

Now we can append the energy of $E(x)$ with all the barriers:
\begin{equation} \label{eq:ipc-energy}
  B(x) = \kappa \sum_{k \in \C} b(d_k(x)),
\end{equation}
where $\kappa$ is a stiffness parameter. Then we are simply facing another unconstrained optimization target $\hat{E}(x) = E(x) + B(x)$. For convenience, we define $\mathcal{B}(x) \overset{\Delta}{=} \nabla^2 B(x)$, and the new global step is:
\begin{equation} 
\label{eq:core}
  \underbrace{(H + \mathcal{B}(x))}_{\hat{H}(x)}\delta x = -(\underbrace{\sum_i w_i G_i\T (G_i x - p_i) + \nabla B(x) }_{\hat{g}(x)}).
\end{equation}
In IPC, continuous collision detection (CCD) is used to find the largest possible step size that does not cause penetration. Starting from the maximal step size, back-traced line search is performed to ensure convergence. We list the complete procedure of the solver in Algorithm ~\ref{alg:pd-ipc}.

\begin{algorithm}
  \caption{PD Solver with IPC Contact}\label{alg:pd-ipc}
  \hspace*{\algorithmicindent} \textbf{Input} $x_t$ \Comment{Equilibrium state at frame $t$ (initial guess)}\\
  \hspace*{\algorithmicindent} \textbf{Output} $x_{t+1}$ \Comment{Equilibrium state at frame $t+1$}
  \begin{algorithmic}
    \State $x \gets x_t$
    \While{not converged}
      \ForAll{$i$} 
      \State $p_i \gets \text{GetProjection}(x, i)$  \Comment{Equation~\ref{eq:local}}
      \EndFor
      \State $\C \gets $ UpdateConstraintSet($x$, $d_0$)
      \State $\hat{H} \gets H + \text{PositiveDefinite}(\mathcal{B}(x, \C))$
      \State $\hatg \gets \sum_i w_i G_i\T (G_i x - p_i)  + \nabla B(x)$
      \State $\dx \gets - \hat{H}\inv \hatg$ \Comment{ Equation~\ref{eq:core}, Section~\ref{sec:opt}}
      \State $\alpha_m \gets $ CCD($x$, $dx$)
      \State $\alpha \gets $ LineSearch($x_t$, $\dx$, $\alpha_m$)
      \State $x \gets x_t + \alpha \cdot \dx$
    \EndWhile
    \State $x_{t+1} \gets x $ 
  \end{algorithmic}
\end{algorithm}

\enlargethispage{15pt}
\section{Optimizing the Global Step} \label{sec:opt}
The fact that $\hat{H}(x)$ in Equation~\ref{eq:core} is no longer constant prevents us from utilizing Cholesky pre-factorization for PD. Re-factorizing the system per iteration is apparently not efficient. Also note that $\hat{H}$ is sparse: given a vertex, its corresponding entries in the Hessian could only be non-zero if another vertex is in the same element (for $H$) or active collision primitive pair (for $\mathcal{B}$). Thus, an iterative method such as conjugate gradient (CG) is a reasonable choice. Furthermore, the constant factors $LL\T = H$ is able to approximate $\hat{H}$ for pre-conditioning the system:
\begin{equation}
  (L^{\sf-1} \hat{H}(x) L^{-{\sf T}}) L\T \dx = -L^{\sf-1} \hat{g}(x).
  \label{eq:cg}
\end{equation}
We use the pre-conditioned CG solver as a baseline method to benchmark our subsequent optimizations.

\enlargethispage{15pt}

\subsection{Localizing Collision via Schur Complement} \label{sec:opt1}
Our first consideration is that we do not want to solve a sparse, large system $\hat{H}$. Inspired by Wang et al. \cite{wang2021local}, we transform this problem into solving a dense, local matrix. Their implementation does not use IPC, but the idea suits our project. The only prerequisite is that we need to manually mark the elements (and their attached vertices) for which we would like to handle collision. This is a reasonable simplification given that collision predominantly happens in specific regions of the face during animation.

\begin{figure}
     \centering
     \begin{subfigure}{0.42\linewidth}
         \centering
         \includegraphics[width=\linewidth]{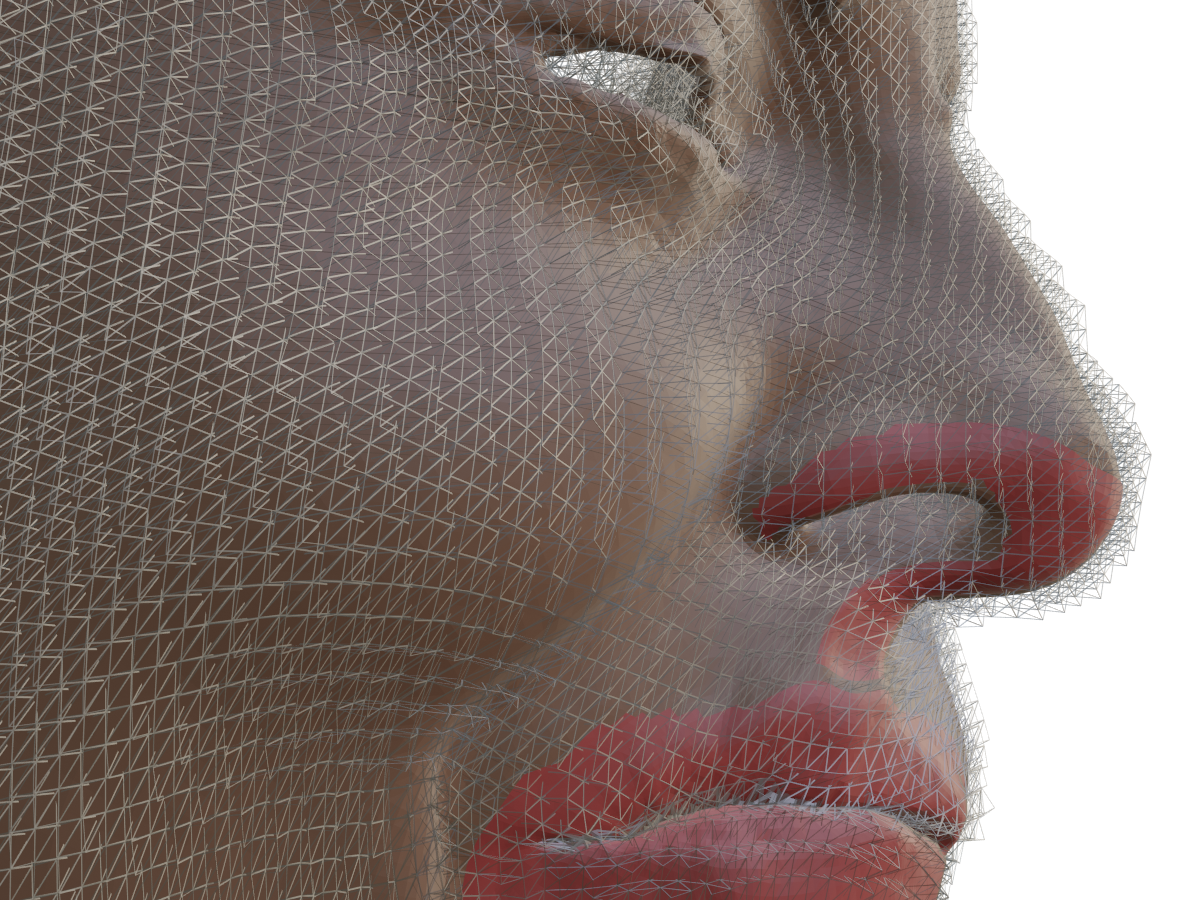}
         \caption{}
         \label{fig:embed}
     \end{subfigure}
   \hfill
     \begin{subfigure}{0.42\linewidth}
         \centering
         \includegraphics[width=\linewidth]{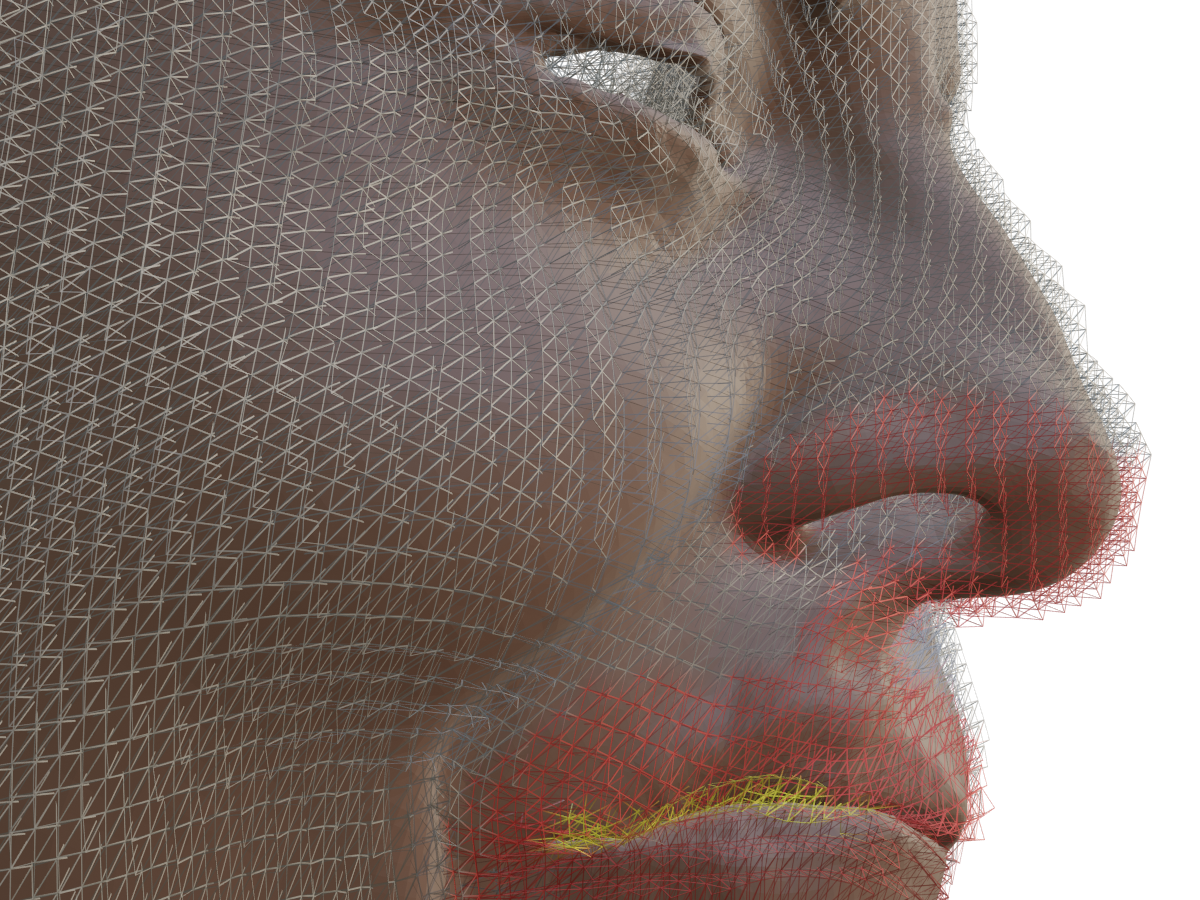}
         \caption{}
         \label{fig:opt}
     \end{subfigure}
     \hfill
    \caption{(a) We use the underlying embedded surface mesh of the face as the proxy for collision response. Specifically, the collision proxy is mainly the mouth region (red). (b) The simulation mesh could be divided into the collision-agnostic part (gray, $n_1$ vertices) and the collision-aware part (red, $n_2$ vertices) that embeds the proxy. During simulation, only a small fraction of the proxy vertices are activated in the constraint set $\C$ and affects the simulation mesh (yellow, $n_c$ vertices). Note how the magnitude of these three hierarchies differ.}
    \label{fig:collision}
    \vspace{-15pt}
\end{figure}

In practice, the simulation mesh $x$ is coarse and does not represent the geometry of the face precisely, which makes it unsuitable for collision detection (see Figure~\ref{fig:embed}). Therefore, we instead employ the embedded surface $s$, which is linearly interpolated as $s = Wx$. Then $s$ serves as the proxy for calculating the primitive pair distance, barrier energy and other quantities in IPC. In this way, we need to map the gradient and Hessian of barriers back to the original space of $x$, which is given by $\nabla B = W\T \nabla B(s)$ and $\mathcal{B} = W\T \nabla^2 B(s) W$.

Since $s$ is only required for the prescribed region, we could split $x$ into two folds: $x_1 \in \mathbb{R}^{3n_1}$ that are free of collision, and $x_2 \in \mathbb{R}^{3n_2}$ that embeds $s$. The idea is shown in Figure~\ref{fig:opt}. We permute the system $\hat{H}$ to cluster these two kinds of vertices:
\begin{equation}
  P \hat{H} P\T = P (H + \mathcal{B}) P\T =
  \begin{pmatrix}
    H_{11} & H_{12} \\
    H_{21} & \hat{H}_{22}
  \end{pmatrix},
\end{equation}
where $P$ is a permutation matrix. Multiplying $P$ and $P\T$ permutes rows and columns respectively. After permutation, $\mathcal{B}$ is only non-zero at the bottom right block that corresponds to $x_2$, thus only $\hat{H}_{22} = H_{22} + \mathcal{B}_{22}$ is non-constant. We pre-factorize the collision-free block $H_{11} = L_1 L_1\T$ before simulation. Now the system could be decomposed as follows, which is an intermediate result of the incomplete Gaussian elimination:
\begin{equation} \label{eq:schur}
  \begin{pmatrix}
  H_{11} & H_{12} \\
  H_{21} & \hat{H}_{22} 
  \end{pmatrix} = 
  \begin{pmatrix}
    L_1 & 0 \\
    H_{21}L_{1}^{\sf-T} & I
  \end{pmatrix}
  \begin{pmatrix}
    I & 0 \\
    0 & \hat{\Sigma}
  \end{pmatrix}
  \begin{pmatrix}
    L_1\T & L_1^{\sf-1} H_{12} \\
    0 & I 
  \end{pmatrix},
\end{equation}
where $\hat{\Sigma}$ is known as the Schur complement of $H_{11}$:
\begin{equation}  \label{eq:system}
  \hat{\Sigma} = \hat{H}_{22} - H_{21} H_{11}^{\sf-1} H_{12} = \mathcal{B}_{22} + \underbrace{H_{22} - H_{21} H_{11}^{\sf-1} H_{12}}_{\Sigma}.
\end{equation}

The significance of this transformation is that we have restricted the influence of IPC to a local matrix $\hat{\Sigma}$. The system in Equation~\ref{eq:schur} can now be solved in 3 steps, simply by solving a lower triangular, a diagonal and an upper triangular system subsequently. Thereafter, the only heavy part is to solve a small, dense system $\hat{\Sigma}^{\sf-1}$ of size $3n_2$, instead of the original sparse $\hat{H}$ of $3n$.

\enlargethispage{15pt}

\subsection{Low-Rank Inverse Update}  \label{sec:opt2}
The previous optimization strategy relies on our manual configuration, which does not consider the actual situation of the collision during the simulation. In practice, it is likely that not all the vertices of $x_2$ are affected by IPC (see Figure~\ref{fig:opt}). Therefore, we further consider how to accelerate solving $\hat{\Sigma}^{\sf-1}$ if only a small fraction of entries has changed during the runtime. Again, we apply the same permutation trick on $\mathcal{B}_{22}$, this time with a matrix $Q \in \mathbb{R}^{3n_2 \times 3n_2}$ that picks out its non-zero entries $\check{\mathcal{B}} \in \mathbb{R}^{3n_c \times 3n_c}$:
\begin{equation}
  Q \mathcal{B}_{22} Q\T =   \begin{pmatrix}
    \check{\mathcal{B}} & 0 \\
    0 & 0 
  \end{pmatrix}
  \; \Leftrightarrow \;
  \mathcal{B}_{22} =
  Q\T 
  \begin{pmatrix}
     \check{\mathcal{B}} & 0 \\
    0 & 0 
  \end{pmatrix}
  Q
  .
\end{equation}
Since the permutation of the remaining zeros is arbitrary, this representation can be simplified as
\begin{equation}
  \mathcal{B}_{22} = 
  \begin{pmatrix}
    \checkQ\T & \cdots 
  \end{pmatrix}
  \begin{pmatrix}
    \checkBeta & 0 \\
    0 & 0 
  \end{pmatrix}
  \begin{pmatrix}
    \checkQ \\
    \vdots 
  \end{pmatrix}
  =
  \checkQ\T \checkBeta \checkQ
  ,
\end{equation}
where $\checkQ \in \mathbb{R}^{3n_c \times 3n_2}$ and $n_c$ is the number of vertices affected by the active barriers.

We pre-compute the explicit inversion of the Schur complement ${\Sigma}^{\sf-1}$ from the PD part as a dense matrix. Thereafter, $\hat{\Sigma}^{\sf-1}$ can be updated efficiently according to the following low-rank update formula (known as Woodbury identity~\cite{hager1989updating}):
\begin{equation}
  \begin{aligned}
    \hat{\Sigma}^{\sf-1} &= (\Sigma + \checkQ\T \checkBeta \checkQ)\inv \\
    & = \Sigma \inv - \Sigma\inv \checkQ\T (\checkBeta\inv + \checkQ\Sigma\inv\checkQ\T)\inv\checkQ\Sigma\inv.
  \end{aligned}
\end{equation}
For each iteration, we need to invert two matrices, namely $\checkBeta\inv$ and $(\checkBeta\inv + \checkQ\Sigma\inv\checkQ\T)\inv$. Once again we have shrunk the scale of the problem from $3n_2$ to $3n_c$. The drawback of this approach is that we need to explicitly invert matrices now, while the previous optimization only requires solving linear systems. Nevertheless, this optimization strategy turns out to be effective because $n_c$ is usually smaller than $n_2$ by magnitudes.
\enlargethispage{10pt}
\section{Experiment}

Our solver is written in Python and runs on GPU (CUDA). Through Python wrapper libraries we employ cuSPARSE for sparse triangular solving and cuSOLVER for dense matrix inversion. Cholesky factorization is performed using CHOLMOD \cite{chen2008cholmod}. As for IPC, we integrate the open-source implementation from the original authors \cite{ipc_toolkit} which only runs on CPU.

\begin{figure} %
    \centering
    \vspace*{-10pt}
    \begin{tabular}{ccc}
        \includegraphics[trim=150 300 150 0, clip, width=0.25\linewidth]{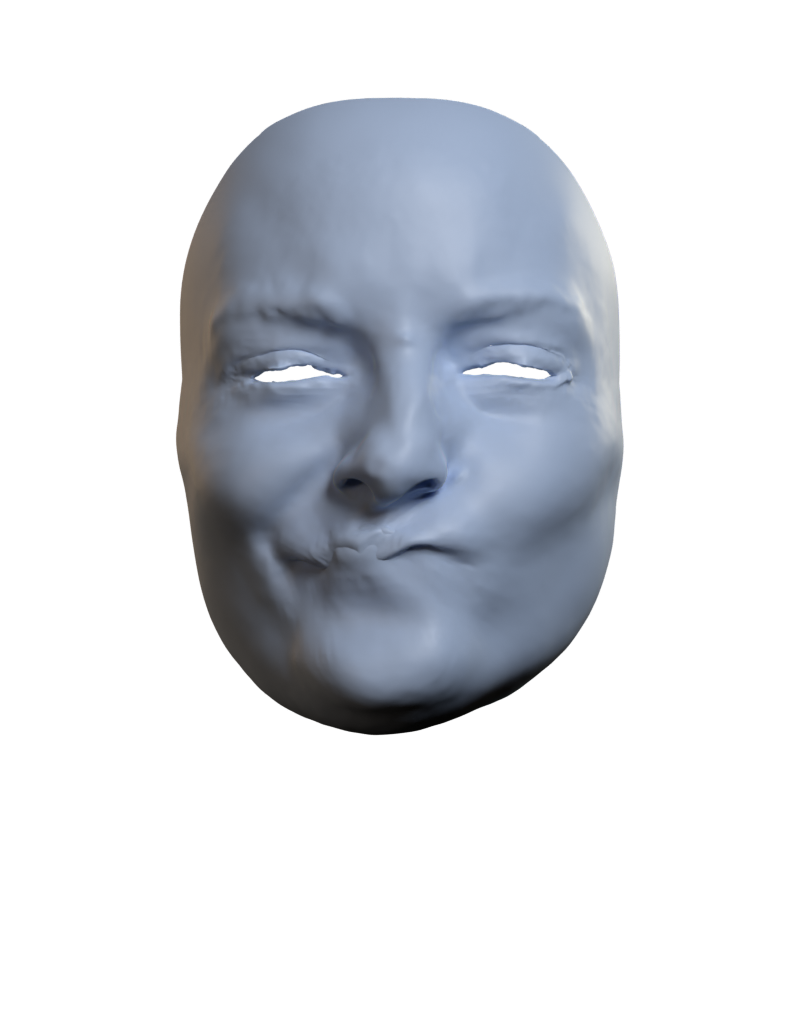} &
        \includegraphics[trim=150 300 150 0, clip, width=0.25\linewidth]{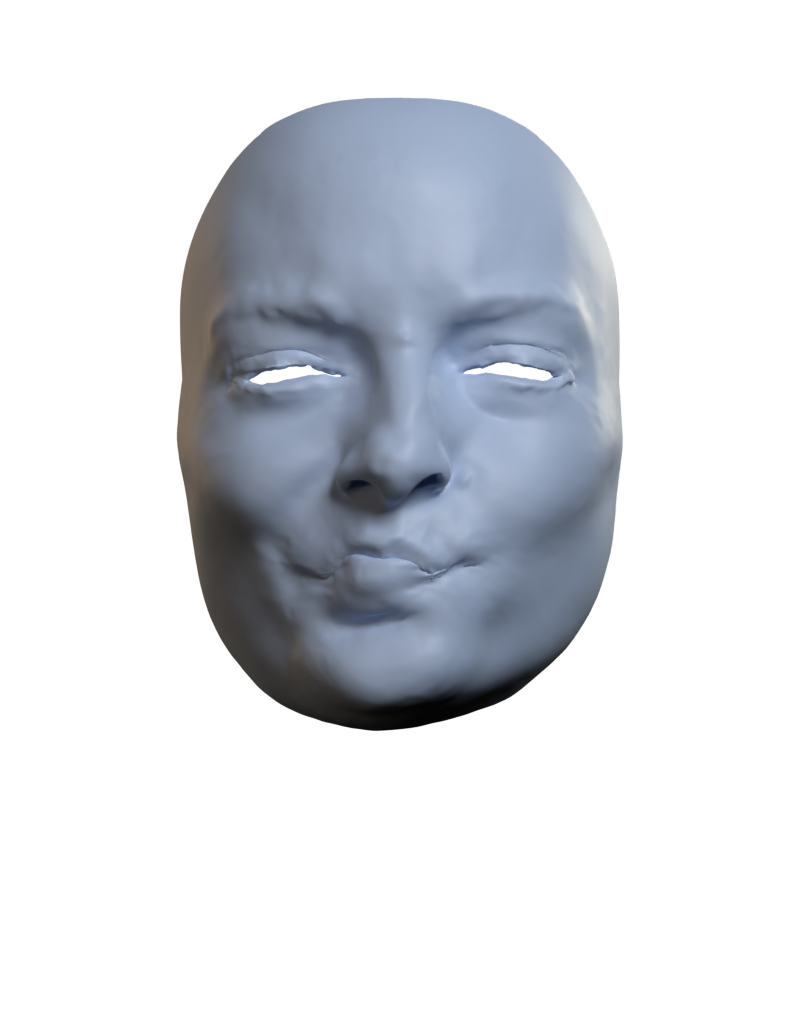} &
        \includegraphics[trim=150 300 150 0, clip, width=0.25\linewidth]{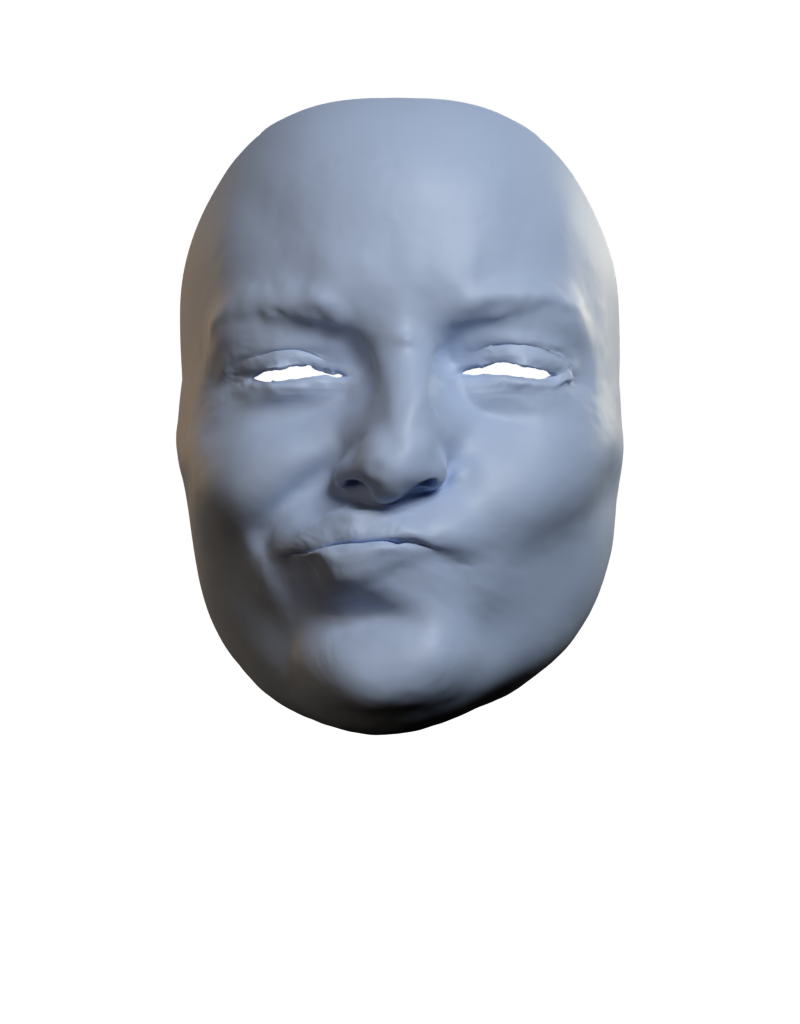} \\
        \includegraphics[trim=200 220 320 0, clip, width=0.2\linewidth]{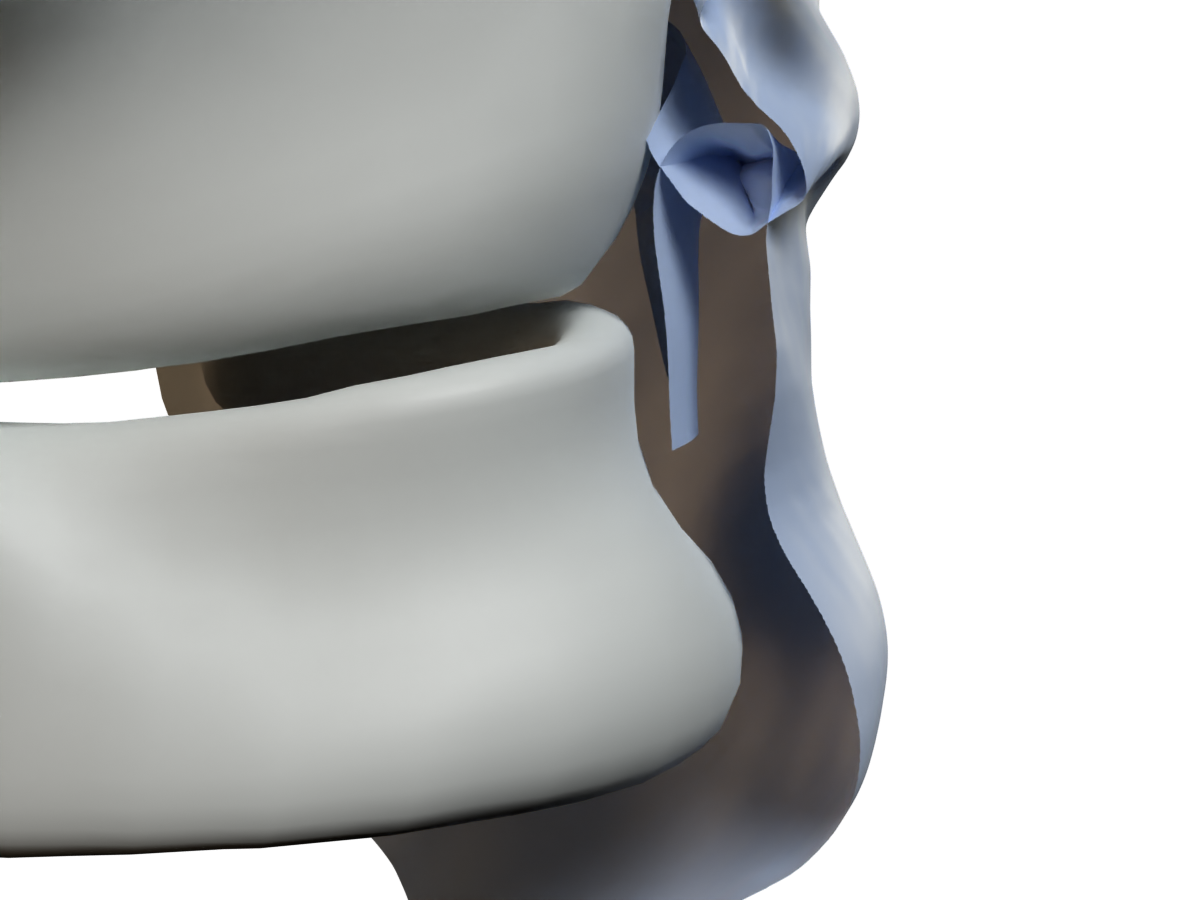} &
        \includegraphics[trim=200 220 320 0, clip, width=0.2\linewidth]{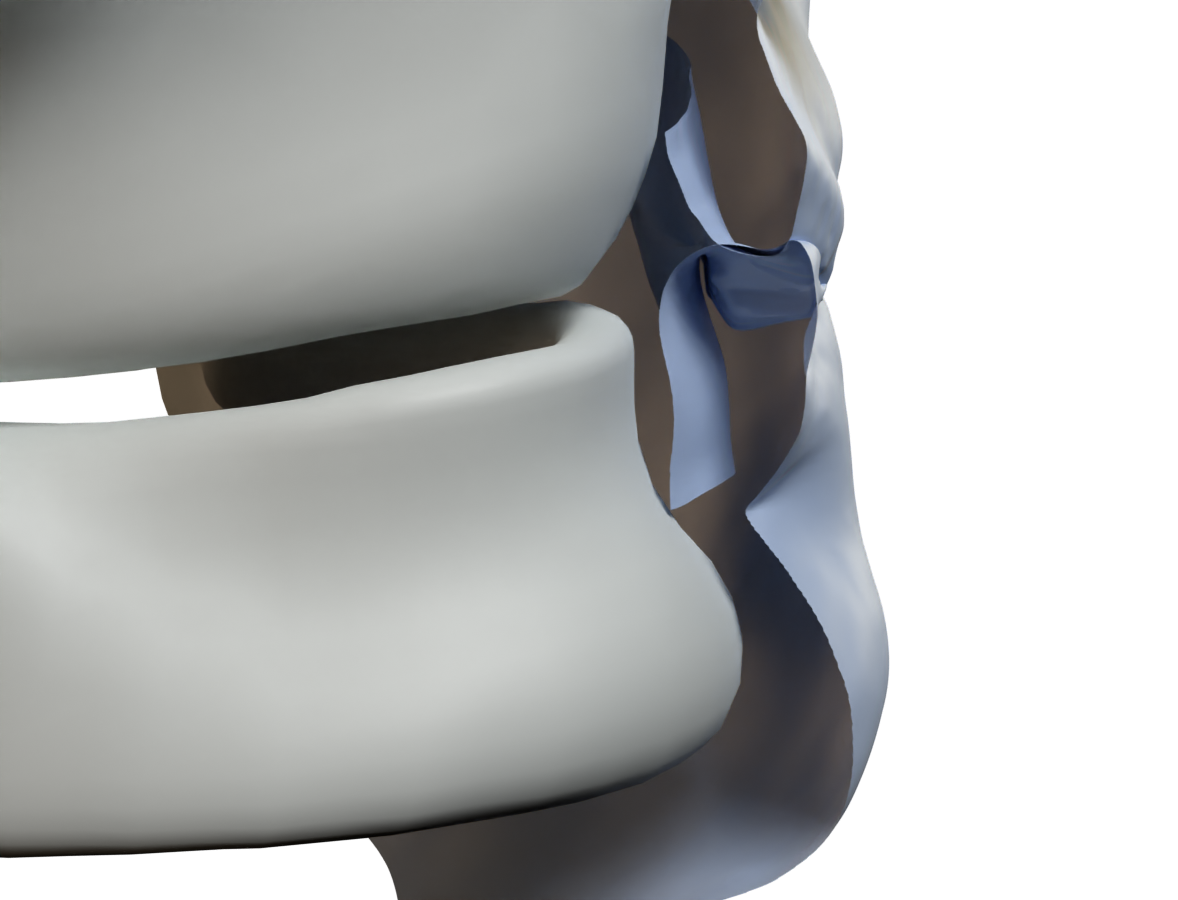} &
        \includegraphics[trim=200 220 320 0, clip, width=0.2\linewidth]{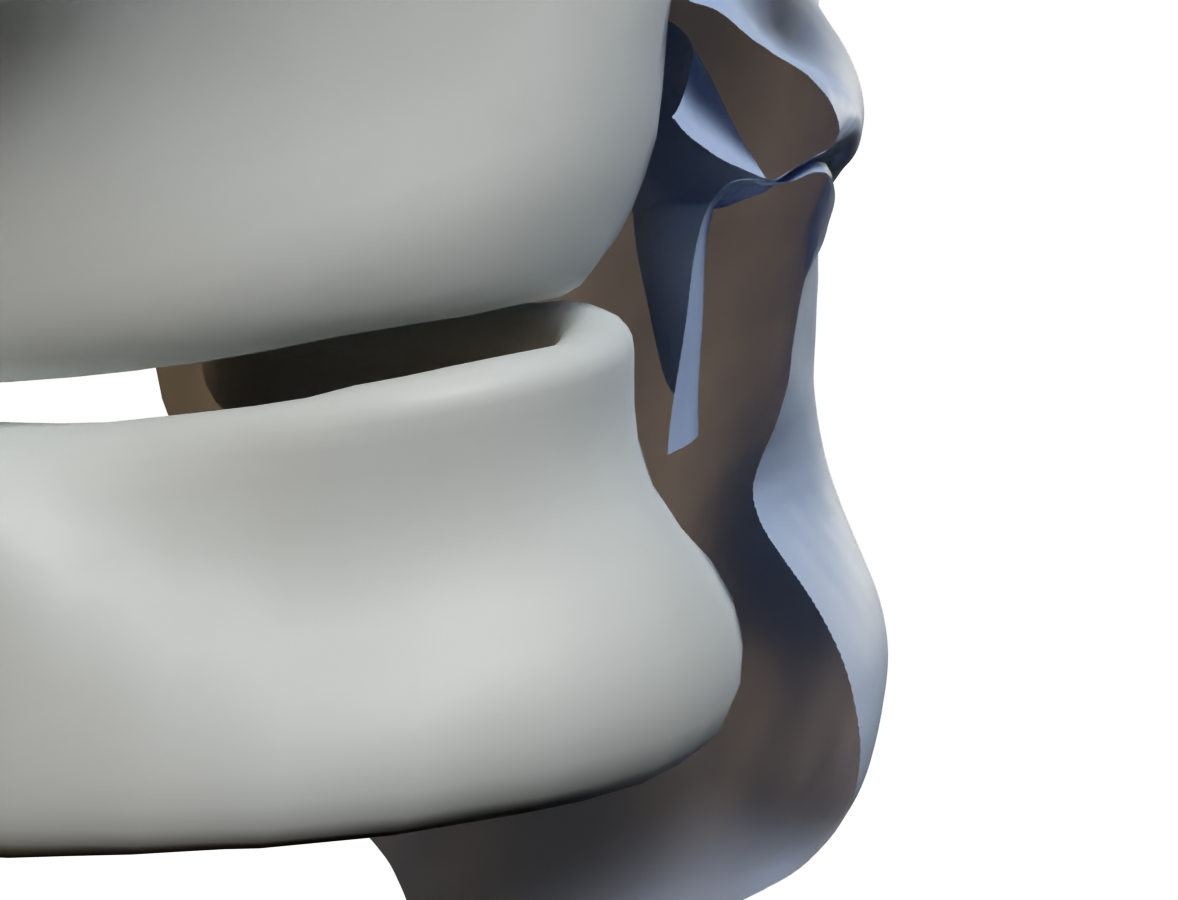} \\
        \small{No Collision} & \small{Standard PD} & \small{IPC}
    \end{tabular}
  \caption{The effect of collision handling. The actuation may lead to severe self-intersection artifacts in extreme cases. Both IPC (ours) and the standard PD can solve the penetration, but IPC results in more natural collision responses.}
  \label{fig:visual}
  \vspace{-5pt}
\end{figure}

We firstly show the effect of collision handling in Figure~\ref{fig:visual}, where we compare three methods: no collision handling; the spring-like repulsion method in the standard PD; ours using IPC. Compared to the collision solution in the standard PD, IPC guarantees the simulation is intersection-free and separates the colliding parts more naturally, thanks to its repulsive contact barrier. 

To benchmark the performance of our optimization for collision, we test our solver on a dataset of three facial animation sequences, with each containing 50 frames. We run the experiment on a PC equipped with an Intel i9-10900KF CPU and a NVIDIA RTX 3060 GPU. The detailed example setup and performance information is listed in Table~\ref{tab:perf}. Note that the prescribed colliding vertices $n_2$ is one magnitude smaller than $n_1$, and during simulation the actual $n_c$ is typically again no more than 10\% of $n_2$, which explains why the optimization works.

Compared to the baseline using conjugate gradient solver, our final optimized version achieves around 3x speedup for the final frame time and 5-7x speedup for the global step. Both optimization strategies roughly contribute equally to the speedup. Moreover, our optimized algorithm has another advantage over the iterative solver. The pre-conditioner in Equation~\ref{eq:cg} becomes less accurate to approximate $\hatH\inv$ if the collision stiffness $\kappa$ becomes larger. In IPC, $\kappa$ is initialized according to the average stiffness of the soft body and adaptively adjusted. If we increase its initial value by 10x, the cost of the conjugate gradient solver would nearly double. However, our method is a direct solver, which does not rely on a pre-conditioner and performs independent of $\kappa$.

Collision handling using IPC is still a relatively expensive solution, and even our optimized version still requires 4x frame time compared to the PD without collision. Besides the fact that the system cannot be easily pre-factorized, the accompanied cost of Hessian computation, CCD and line search with IPC is also significant. In our final optimized implementation, they almost occupy half of the total frame time and would hinder the overall benefit for further optimizing the global step only.

\begin{table}[!hbt]
\caption{The average per-frame performance data for three test cases. The meaning of the steps are: \textbf{IPC}: constructing the IPC constraint set, computing the gradient and Hessian (with positive-definite regularization) of the barrier function; \textbf{G-Step}: solving the global step; \textbf{CCD}: continuous collision detection; \textbf{LS}: line search;  \textbf{Misc}: the remaining items, including the local step, matrix assembly, data transfer between CPU/GPU, etc. The time is measured in milliseconds. For each test case, we experiment with four configurations: $\circ$: without collision; $\triangle$: IPC with pre-conditioned conjugate gradient solver; $\square$: optimization as in Section~\ref{sec:opt1}; $\diamondsuit$: additional optimization as in Section~\ref{sec:opt2}. 
\label{tab:perf}
}
\vspace{-10pt}
\scalebox{0.80}{
\begin{tabular}{c|cc|ccccc|cc}
\hline
\multicolumn{1}{c|}{\textbf{}}  & \textbf{$n_1$}         & \textbf{$n_2$}        & \textbf{IPC} & \textbf{G-Step} & \textbf{CCD} & \textbf{LS} & \textbf{Misc} & \textbf{Total} &              \\ \hline
\multirow{4}{*}{\texttt{Case 1}} & \multirow{4}{*}{51.5k} & \multirow{4}{*}{3.7k} & ---          & 418             & ---          & ---         & 173           & 591            & $\circ$        \\
                                &                        &                       & 451          & 6059            & 348          & 361         & 171           & 7390           & $\triangle$    \\
                                &                        &                       & 472          & 2923            & 401          & 378         & 172           & 4346           & $\square$      \\
                                &                        &                       & 470          & \textbf{1400}            & 396          & 355         & 166           & \textbf{2787}           & $\diamondsuit$ \\ \hline
\multirow{4}{*}{\texttt{Case 2}}  & \multirow{4}{*}{53.0k} & \multirow{4}{*}{3.9k} & ---          & 407             & ---          & ---         & 182           & 589            & $\circ$        \\
                                &                        &                       & 412          & 6861            & 448          & 493         & 228           & 8442           & $\triangle$    \\
                                &                        &                       & 428          & 2978            & 463          & 405         & 154           & 4427           & $\square$      \\
                                &                        &                       & 426          & \textbf{976}             & 416          & 399         & 158           & \textbf{2375}           & $\diamondsuit$ \\ \hline
\multirow{4}{*}{\texttt{Case 3}} & \multirow{4}{*}{59.0k} & \multirow{4}{*}{4.1k} & ---          & 436             & ---          & ---         & 191           & 627            & $\circ$        \\
                                &                        &                       & 419          & 6101            & 274          & 290         & 156           & 7240           & $\triangle$    \\
                                &                        &                       & 434          & 3163            & 275          & 334         & 177           & 4383           & $\square$      \\
                                &                        &                       & 436          & \textbf{1013}            & 274          & 325         & 170           & \textbf{2217}           & $\diamondsuit$ \\ \hline
\end{tabular}
}

\vspace{-12pt}
\end{table}

\enlargethispage{15pt}
\section{Conclusions}
Our solver is able to animate human face and is aware of collision. Currently, our optimization for collision handling still relies on manually prescribing the region of interest, and it would be beneficial to explore fully automatic solutions. Moreover, exploiting collision detection on GPU and faster CCD approximation may help address our performance bottleneck caused by IPC. Also, the idea of our optimization might find application in other problems with a similar setting, which involves solving a system with a fixed and varying part. Finally, though our work is implemented for quasi-static actuated face simulation, it should be straightforward to extend it to other types of soft bodies and dynamics simulation.

\bibliographystyle{ACM-Reference-Format}
\bibliography{reference}

\end{document}